\documentclass[twocolumn,aps,superscriptaddress,showpacs,nofootinbib]{revtex4-1}
\usepackage[english]{babel}
\usepackage{epsfig}
\usepackage{graphicx}
\usepackage{color}
\usepackage{amsfonts,amsmath,amssymb}
\usepackage{enumerate}

\newcommand{\nc}{\newcommand}
\def\cal{\mathcal}
\def\rm{\mathrm}

\nc{\eq}[1]{Eq.~(\ref{#1})}
\nc{\eqs}[1]{Eqs.~(\ref{#1})}
\nc{\beq}{\begin{equation}}
\nc{\eeq}{\end{equation}}
\nc{\bal}{\begin{alignedat}}
\nc{\eal}{\end{alignedat}}
\nc{\bea}{\begin{eqnarray}}
\nc{\eea}{\end{eqnarray}}
\nc{\bit}{\begin{itemize}}
\nc{\eit}{\end{itemize}}
\nc{\benu}{\begin{enumerate}}
\nc{\eenu}{\end{enumerate}}
\nc{\bdes}{\begin{description}}
\nc{\edes}{\end{description}}
\nc{\nn}{\nonumber}

\nc{\sub}[1]{_{\rm{#1}}}
\nc{\ssub}[1]{_{_\rm{#1}}}
\nc{\super}[1]{^{\rm{#1}}}
\nc{\ssuper}[1]{^{^\rm{#1}}}

\nc{\slashed}[1]{{#1}\hspace{-1.8mm}/}

\nc{\pare}[1]{\left( #1 \right)}
\nc{\sqpare}[1]{\left[ #1 \right]}
\nc{\ang}[1]{\langle #1 \rangle}
\nc{\abs}[1]{\left| #1 \right|}

\def\cc{\rm{c.c.}}

\def\g5{\gamma_{5}}

\def\GeV{\: \rm{GeV}}
\def\TeV{\: \rm{TeV}}

\def \cm{\: \rm{cm}}

\def\a{\alpha}

\def\g{\gamma}
\def\d{\delta}
\def\e{\epsilon}
\def\z{\zeta}
\def\h{\eta}
\def\th{\theta}
\def\i{\iota}
\def\k{\kappa}
\def\l{\lambda}
\def\m{\mu}
\def\n{\nu}

\def\p{\pi}

\def\s{\sigma}

\def\f{\phi}

\def\ps{\psi}

\def\vs{\varsigma}

\def\G{\Gamma}

\def\F{\Phi}

\def\Ps{\Psi}
\def\W{\Omega}

\def\Fh{\hat{\F}}
\def\fh{\hat{\f}}
\def\psh{\hat{\ps}}

\def\kh{\hat{\k}}

\def\msb{\widetilde{m}}

\begin{document}

\title{Pangenesis in a Baryon-Symmetric Universe: \\ Dark and Visible Matter via the Affleck-Dine Mechanism.}

\author{Nicole F. Bell}
\affiliation{ARC Centre of Excellence for Particle Physics at the Terascale, School of Physics, The University of Melbourne, Victoria 3010, Australia}
\author{Kalliopi Petraki}
\email{Corresponding author, kpetraki@unimelb.edu.au}
\affiliation{ARC Centre of Excellence for Particle Physics at the Terascale, School of Physics, The University of Melbourne, Victoria 3010, Australia}
\author{Ian M. Shoemaker} 
\affiliation{Theoretical Division T-2, Los Alamos National Laboratory, Los Alamos, NM 87545, USA}
\author{Raymond R. Volkas} 
\affiliation{ARC Centre of Excellence for Particle Physics at the Terascale, School of Physics, The University of Melbourne, Victoria 3010, Australia}

\begin{abstract}
The similarity of the visible and dark matter abundances indicates that they may originate via the same mechanism. If both the dark and the visible matter are charged under a generalized baryon number which remains always conserved, then the asymmetry of the visible sector may be compensated by an asymmetry in the dark sector. We show how the separation of baryonic and antibaryonic charge can originate in the vacuum, via the Affleck-Dine mechanism, due to the breaking of a symmetry orthogonal to the baryon number. Symmetry restoration in the current epoch guarantees the individual stability of the two sectors.
\end{abstract}


\maketitle

\section{Introduction}
\label{sec:intro} 

The energy content of our universe today is a fossil of the processes that took place at its early stages. 
About 5\% of our universe consists of visible matter, while 23\%  consists of dark matter (DM).
The relic abundance of either of these components requires physics beyond what has currently been established.

The density of visible matter today is determined by a baryon asymmetry, an excess of particles over antiparticles, that was established in the early universe. It may be quantified by the current baryon-to-entropy ratio $\h(B) \equiv [n(B) - n(\bar{B})]/s  \simeq 10^{-10}$.  This asymmetry persists because the (unknown) baryon number violating processes that gave rise to it have become ineffective in the low-energy late universe.

It is possible that the DM density today is also determined by a DM particle/antiparticle asymmetry. Indeed, the similar densities of the visible and the dark matter suggest a common origin. If the abundances resulted from unrelated mechanisms, they would depend on different parameters, and be expected, generically, to vary greatly. In this paper, we outline a new mechanism which relates the relic densities of the two matter components of the universe.
We shall thus call it \emph{pangenesis}\footnote{From Greek ``$\p\a\n$'' = all and ``$\g\acute{\e}\n\e\s\i\vs$'' = birth, here meaning the origin of all matter.}. In this mechanism, the two sectors are charged under a generalisation of baryon number, which remains conserved, while they develop compensating asymmetries. The possibility of a baryon-symmetric universe has been studied in 
Refs.~\cite{Dodelson:1989ii,Dodelson:1989cq,Dodelson:1990ge,Kuzmin:1996he,Oaknin:2003uv,Kitano:2004sv,Agashe:2004bm,Farrar:2005zd,Gu:2007cw,Gu:2009yy,An:2009vq,Davoudiasl:2010am,Gu:2010ft}.
Here we employ a different technique, the Affleck-Dine (AD) mechanism~\cite{Affleck:1984fy,Dine:1995kz}, for the separation of the baryon--antibaryon number, which befits a different cosmology.

The separation of baryon and antibaryon number into two sectors necessitates a particular symmetry structure. Each sector is stabilised at low energies by an unbroken abelian symmetry: this can be the anomaly-free $(B-L)_1$ for the visible sector, and $B_2$ for the dark-sector. We define the linearly independent charges  $B-L$ and $X$
\beq
\bal{2}
B-L &\equiv (B-L)_1 - B_2,  \\
X   &\equiv (B-L)_1 + B_2,
\eal
\label{eq:B-L,X def}
\eeq
with symmetry generators $T_{_{B-L}}, T_{_X}$ respectively. $B-L$ remains always conserved, while $X$ breaks at some high-energy scale. If the $X$-violating interactions occur out of thermal equilibrium and while CP violation is in effect, a net $X$ charge will be created. A non-zero $X$ and a vanishing $B-L$ allots a net charge 
\beq
\h\pare{(B-L)_1} = \h\pare{B_2} = \h\pare{X}/2
\label{eq:asym rel}
\eeq 
to the two sectors, thus relating the number densities of the visible baryonic and dark antibaryonic matter (provided that any fields carrying both $(B-L)_1$ and $B_2$ have negligible relic density, due to e.g. decay, or thermal suppression). In the scenario we present here, a net $X$ number arises via the AD mechanism.

In the AD mechanism, the oscillations of a scalar condensate can give rise to coherent production of a net charge if a symmetry is explicitly broken. This process is not coupled to the thermal history of the universe, and can successfully generate an asymmetry even if the energy scale of the symmetry violation was never accessible by the thermal bath. Condensates arise quite generically in cosmological models, and in particular in the context of supersymmetry, which is a leading candidate for physics beyond the standard model (SM). The AD scenario is thus regarded as one of the most plausible asymmetry-generation mechanisms~\cite{Dine:2003ax,Enqvist:2003gh}.
Here we show how in extensions of the minimal supersymmetric standard model (MSSM), it can separate the baryonic and antibaryonic charge.

Other scenarios in which both the dark and the visible matter arise via the AD mechanism have been studied in
Refs.~\cite{Kusenko:1997si,Thomas:1995ze,Enqvist:1998en,Fujii:2002kr,Fujii:2002aj,Roszkowski:2006kw,McDonald:2006if,Kitano:2008tk,Shoemaker:2009kg,Higashi:2011qq,Doddato:2011fz}.
Models in which the DM shares the asymmetry of the visible matter have also been developed in 
Refs.~
\cite{Nussinov:1985xr,Barr:1990ca,Barr:1991qn,Kaplan:1991ah,Berezhiani:2000gw,Foot:2003jt,Foot:2004pq,Hooper:2004dc,Cosme:2005sb,Suematsu:2005kp,Suematsu:2005zc,Gudnason:2006ug,Gudnason:2006yj,Banks:2006xr,Dutta:2006pt,Berezhiani:2008zza,Khlopov:2008ty,Ryttov:2008xe,Foadi:2008qv,Kaplan:2009ag,Kribs:2009fy,Cohen:2009fz,Cai:2009ia,Frandsen:2009mi,Gu:2010yf,Dulaney:2010dj,Cohen:2010kn,Shelton:2010ta,Haba:2010bm,Buckley:2010ui,Chun:2010hz,Blennow:2010qp,McDonald:2010rn,Hall:2010jx,Allahverdi:2010rh,Dutta:2010va,Falkowski:2011xh,Haba:2011uz,Chun:2011cc,Kang:2011wb,Graesser:2011wi,Frandsen:2011kt,Kaplan:2011yj}.
They differ from pangenesis in the symmetry structure and/or the asymmetry-generation mechanism.

\section{The mechanism}
\label{sec:mech}

The effectiveness of the AD mechanism relies on the existence of flat directions (FDs) in the scalar potential~\cite{Affleck:1984fy,Dine:1995kz,Dine:2003ax,Enqvist:2003gh}.
The continuous degeneracy of the vacuum along a FD allows the scalar field to pick up a large vacuum expectation value (VEV) in the early universe. A large VEV amplifies the effect of small symmetry-violating terms that arise at higher orders along the FD, and can lead to an appreciable asymmetry generation. FDs occur generically in susy models.

Pangenesis occurs along FDs with $D_{_{B-L}} \equiv \f^\dag T_{_{B-L}} \f = 0$ and $D_{_X} \equiv \f^\dag T_{_X} \f \ne 0$, where $\f$ parametrizes the FD. In models with gauged $B-L$ and low-scale $B-L$ breaking, the vanishing of $D_{_{B-L}}$ is warranted by D-flatness. If $B-L$ is not gauged, explicit $(B-L)$-breaking is generically expected by non-renormalisable operators. However, such terms are inoperative along directions which do not carry $B-L$, or directions which break $B-L$ spontaneously while maintaining $D_{_{B-L}} = 0$. Pangenesis can still be realized along such directions. Here, we will present in some detail the pangenesis mechanism along field directions with $T_{_{B-L}} \f=0$ and $D_{_X} \ne 0$. We shall explore the other possibilities elsewhere.

\section{A model of Pangenesis}

We introduce three SM gauge-singlet chiral superfields, $\F_j = (\f_j, \ps_j, F_j)$ where $j=0,1,2$, and their vector-like partners  $\Fh_j = (\fh_j, \psh_j, \hat{F}_j)$, with the baryonic charge assignments presented in Table~\ref{tab:charges}.
\begin{table}[t]
\begin{center}
\begin{tabular}{c|cccccc}
\hline \hline
          & $\F_1$ & $\F_2$ & $\F_0$ & $\Fh_1$ & $\Fh_2$ & $\Fh_0$  \\ \hline  %\hline 
$(B-L)_1$ &  1     &  0     &  -1    &  -1     &    0    &  1       \\ %\hline
$B_2$     &  0     &  1     &  -1    &   0     &   -1    &  1       \\ %\hline
$B-L$     &  1     & -1     &   0    &  -1     &    1    &  0       \\ %\hline
$X$       &  1     &  1     &  -2    &  -1     &   -1    &  2       \\ 
\hline\hline
\end{tabular}
\end{center}
\caption{Baryoleptonic charge assignments.}
\label{tab:charges}
\end{table}
At the renormalizable level, we require that both $B-L$ and $X$ are conserved. The superpotential terms involving the new fields are
\beq
\d W_\rm{r} = 
\k \F_0 \F_1 \F_2  + \kh \Fh_0 \Fh_1 \Fh_2  + \sum_{j=0}^2 \m_j \F_j \Fh_j.
\label{eq:fee-fau-fum}
\eeq
The scalar components $\f_j, \fh_j^*$ for each $j$ mass-mix (due to  susy-breaking mass contributions) to produce six complex scalar mass eigenstates. The fermionic components form three Dirac fermions $\Ps_j^\rm{T} = \pare{{\ps_j}_\a, \ \psh_j^{\dag \dot{\a}}}$ of mass $\m_j$.
The $\F_1, \Fh_1$ supermultiplets have $(B-L)_1 \ne 0, B_2=0$, belong to the visible sector and couple to the MSSM. $\F_2, \Fh_2$ have $(B-L)_1=0, B_2 \ne 0$, and belong to the dark sector. $\F_0, \Fh_0$ are connector fields, carrying both $(B-L)_1$ and $B_2$. Conservation of $B-L$ and $X$ forbids renormalizable couplings of $\F_0, \Fh_0$ to either sector.
We discuss the visible and dark sector couplings in Secs.~\ref{sec:vis}, \ref{sec:dark}.
The masses are such that $\F_0, \Fh_0$ decay into visible and dark sector fields, $\F_1, \Fh_1$ decay into MSSM particles, and $\F_2, \Fh_2$ possibly decay into other dark-sector particles.

The directions in the scalar potential parametrized by $\f_0$ and $\fh_0$ (with $\f_{1,2} \! = \! \fh_{1,2} = 0$) are flat, up to terms of positive mass dimension. Such terms are also generated by susy breaking, and do not destroy the flatness of the potential, since they are suppressed by factors of the mass parameter over the field VEV, where the latter can be quite large in the early universe. 
The $\f_0, \fh_0$ directions carry $X$ number. In pangenesis, a $\f_0, \fh_0$ condensate results from a 2nd-order phase transition in the early universe, and develops a net $X$-charge due to $X$-violating terms suppressed by a large scale. The subsequent $X$-conserving decay or evaporation of the condensate transfers the $X$ charge into the visible and dark sectors, each of which acquires a baryonic asymmetry as per \eq{eq:asym rel}.

Since $\F_0, \Fh_0$ do not have renormalizable couplings to the MSSM fields, the MSSM flat directions can in principle be operational simultaneously. 
If AD baryogenesis were implemented successfully along a $(B-L)_1$-charged FD of the MSSM, the correlation between the visible and dark sector asymmetries would be destroyed. Gauging $B-L$ precludes this possibility. Alternatively, assuming minimal K\"{a}hler potential for the MSSM fields prevents the latter from acquiring large VEVs.

The potential arising from \eq{eq:fee-fau-fum} has two more pairs of FDs, parametrized by $\f_1, \fh_1$ and by $\f_2, \fh_2$. These FDs cannot be operational simultaneously with $\f_0,\fh_0$ due to the F-terms in the superpotential. The singling out of the $\f_0,\fh_0$ FDs in the early universe can be similarly argued on the basis of choosing a minimal K\"{a}hler potential along the $\f_1,\fh_1,\f_2,\fh_2$ directions, and/or gauging $B-L$.

The FDs of susy models are typically lifted by non-renormalizable terms in the superpotential, and by susy breaking. 
At the non-renormalizable level, we impose no global symmetries. The superpotential terms which contribute to lowest order across the $\f_0, \fh_0$ plane, are
\beq
\d W_\rm{nr} \supset \frac{1}{M} \: \F_j \F_0^k \Fh_0^{3-k}, \ \frac{1}{M} \: \Fh_j \F_0^k \Fh_0^{3-k}
\label{eq:Wnr}
\eeq
with $j=0,1,2$ and $k=0,1,2,3$. $M$ is the scale of new physics, usually assumed to be close to or at the Planck scale. 
Equation~\eqref{eq:Wnr} includes $X$-violating contributions. 
Susy breaking is expected to arise from the hidden sector, and due to the finite energy density of the universe.
Including susy and susy-breaking terms, the scalar potential on the $\f_0,\fh_0$ flat manifold is 
\beq
\bal{2}
V_\rm{AD} 
&= \! \sqpare{m_0^2(T) - c H^2} |\f_0|^2  + \sqpare{\hat{m}_0^2(T)  -  \hat{c} H^2} |\fh_0|^2   \\ 
+&  b \msb \mu_0 \f_0 \fh_0  +  \frac{\m_0}{M} \sum_{k=0}^3 \pare{\z_k \f_0 + \hat{\z}_k \fh_0} \f_0^{* k}  \fh_0^{* {3-k}}  \\
+& \sum_{k=0}^4 \frac{(A_k \msb + a_k H + f_k H^2/M) \l_k}{M} \f_0^k \fh_0^{4-k}  \\
+& \sum_{k=0}^3 \sum_{l=0}^{3-k} \frac{\l^2_{kl}}{M^2} \pare{\f_0^* \fh_0}^{3-k-l}|\f_0|^{2k}|\fh_0|^{2l} + \cc, 
\eal
\label{eq:VFD}
\eeq
where $H$ is the Hubble constant.

In the above, $m_0^2(T) \simeq \m_0^2 + \msb^2 + \k^2 T^2$, and similarly for $\hat{m}_0^2(T)$.
The hidden-sector susy breaking generates the $\msb$ terms.
The thermal masses arise from the Yukawa couplings (see \eq{eq:fee-fau-fum}) with the dilute thermal bath of temperature $T~\approx~\sqpare{T_R^2 H(t) M_P}^{1/4}$, present 
after inflation but before reheating is completed at temperature $T_R$. Moreover, the vacuum energy density induces mass-squared contributions for $\f_0,\fh_0$ which can be negative, of order $-cH^2$ with $c \sim \mathcal{O}(1)$, if the coupling of the scalars to the inflaton is non-minimal. The negative mass terms dominate in the early universe, and ensure that the fields along the FDs acquire large VEVs.

The susy-breaking $A$-terms induced by the hidden sector and the vacuum energy density are of order $A\msb \: \d W_\rm{nr}$ and $(a H + f H^2/M) \: \d W_\rm{nr}$ respectively, where $A,a,f$ are complex coefficients of $\cal{O}(1)$. These terms are typically responsible for the generation of a net charge in the condensate. They violate explicitly the $U(1)$ symmetry (here, the $X$ number) carried by the condensate fields at low energies. The phase differences between the various contributions are sources of CP violation.

The sextic terms of Eq.\eqref{eq:VFD} result from the non-renormalizable terms in Eq.\eqref{eq:Wnr}, and stabilize the potential at high VEVs. 
They have relative phases, which can induce CP violation, even immediately after inflation. This is typical in the multifield AD mechanism, and can enhance the asymmetry generated in the condensate~\cite{Senami:2002kn,Kamada:2008sv,Enqvist:2003pb}, which may be otherwise suppressed by thermal effects~\cite{Allahverdi:2000zd,Anisimov:2000wx,Anisimov:2001dp}.

Following previous analyses~\cite{Senami:2002kn,Kamada:2008sv}, we estimate the $X$-charge-to-entropy ratio generated
\beq
\eta(X) \sim 10^{-10} \pare{ \frac{\sin \d}{\l} } \pare{\frac{T_R}{10^9 \GeV}} \frac{M}{M_P},
\label{eq:eta X}
\eeq
where $\d$ is the effective CP-violating phase, and $\l$ stands for the scale of the smallest coupling in \eq{eq:VFD}.
This asymmetry is generated during inflaton oscillations, when the energy density of the universe redshifts as $R^{-3}$. 
The AD condensate also oscillates coherently, and redshifts at the same rate.  
$\h(X)$ will remain frozen, if the condensate does not decay or evaporate into radiation before reheating.

Evaporation occurs due to elastic scattering of relativistic particles off the condensate.
The scattering cross section for each scalar $\f_j,\fh_j$ is  $\s_\rm{s} = \k^4/ 32 \p m_0 E_j$, 
where we took $\kh \sim \k$. The mass $m_0$ of $\f_0, \fh_0$ is dominated by the thermal contribution $\propto \k T$ till quite later after reheating. 
The scattering of fermions on the condensate is subdominant. The condensate does not evaporate before reheating  if $\ \k, \kh <  10^{-2} \pare{T_R/10^9 \GeV}^{1/3}$.

The $\f_0, \fh_0$ particles decay, partitioning $X$ into the two sectors.  The decay is dominantly into fermions, with rate $\G_0 \simeq \k^2 m_0/16 \p$.
This places their decay temperature, $T_0 \approx 10^6 \GeV \pare{\k/10^{-3}} \pare{m_0/ 1 \TeV}^{1/2}$, comfortably after reheating and before the freeze-out of sphaleron transitions. 
The latter is important if the condensate couples to the MSSM fields only leptonically, as we shall see is the case with the model presented here. 
The AD condensate may also fragment into $Q$-balls~\cite{Kusenko:1997si}. While their formation and stability is model dependent, any $Q$-balls formed can be arranged to decay before the sphaleron and the LSP freeze-out by choosing $\mu_0$ (which increases the curvature of the potential along the FDs) to be sufficiently large.

The $\Ps_0$ fermions are $R$-parity odd particles. 
If $\m_0$ is of the order of the susy-breaking masses or higher, they can decay at rate $\G(\Ps_0) \simeq \k^2 \m_0/4 \p$, before the sphaleron freeze-out. 
For lower values of $\m_0$, the $\Ps_0$ fermions can decay provided that $R$-parity is broken (see Sec.~\ref{sec:dark}).

\subsection{The visible sector}
\label{sec:vis}

The only gauge-invariant renormalizable coupling of $\F_1, \Fh_1$ to MSSM fields is
\beq
\d W_1 = y_1 \F_1 L H_u .
\label{eq:LHu}
\eeq
$\F_1$ couples as a right-handed neutrino superfield, but one linear combination of $\nu$ and $\Ps_1$ remains massless.
After the electroweak phase transition, the mixing of $\Ps_1$ with $\nu$ 
is $\th_\n^2 \approx y_1^2 v^2/2 \m_1^2$, which complies with the experimental bounds, currently standing at $\th_\n^2 \lesssim 10^{-4}$ for masses of a few GeV or higher~\cite{Smirnov:2006bu}, if $y_1 < 10^{-3} \pare{\mu_1/10 \GeV}$.

The $\F_1$ fields communicate the $(B-L)_1$ asymmetry to the MSSM, as a net lepton number. 
Sphalerons reprocess the latter into a baryon number $B_1 = \a (B-L)_1$, with $\a \approx 0.3$, 
provided that the $\F_1$ components decay into or equilibrate with the visible sector before the sphaleron freeze-out. 
The $\f_1, \fh_1$ scalars, and also the $\Ps_1$ fermion for $\m_1 \gtrsim \msb$, decay in time if $y_1 \gtrsim 10^{-7} $. In addition, the $\Ps_1$ particles thermalize for $y_1 \gtrsim 10^{-7} \pare{ \m_1/ 10 \GeV}^{1/2} $~\cite{Smirnov:2006bu}.

\subsection{The dark sector} 
\label{sec:dark}

Equation \eqref{eq:asym rel} relates the relic number densities of the visible and the dark sectors only if the symmetric part of DM is efficiently annihilated. 
If DM thermalizes, the DM annihilation cross section needs to exceed the canonical value for symmetric thermal DM %, albeit only 
by a factor of a few~\cite{Graesser:2011wi}. The prospects for annihilation into SM states are thus severely limited, as such an interaction would have to be comparable or stronger than the weak force~\cite{Bai:2010hh,Goodman:2010ku}. In particular, a gauged $B-L$ cannot efficiently annihilate the symmetric part of DM.

This suggests the existence of a dark interaction that compels DM to annihilate primarily into dark-sector radiation.
Constraints on the relativistic energy density present at the time of big-bang nucleosynthesis are satisfied 
if the dark sector decouples early and evolves at a lower temperature than the visible sector, due to larger entropy release and reheating in the latter.

If the annihilation of DM proceeds via a dark Yukawa coupling, then DM need not carry any gauge charges, and can simply be the $\Ps_2$ fermion. If the DM annihilates via a vector interaction, though, some other $B_2$-charged particle has to play the role of DM. In either case, the minimal extension of the MSSM described by \eqs{eq:fee-fau-fum} and \eqref{eq:LHu} has to be further augmented to accommodate the above. Here we do not introduce any specific dark-sector couplings, as the possibilities are unlimited. We will only make some model-independent remarks.

The prediction of the DM mass is generic in models in which the asymmetric part dominates. In pangenesis
\beq
m_\rm{DM} = q_{_\rm{DM}} \frac{\W_\rm{DM}}{\W_\rm{VM}}  \frac{ \h(B_1) }{ \h(B_2) }    m_p \sim 5 \GeV,
\label{eq:DM mass}
\eeq
where $\W_\rm{DM} \! \simeq \! 0.23$ and  $\W_\rm{VM} \! \simeq \! 0.046$~\cite{Komatsu:2008hk}. The above value can vary by a factor of a few in different implementations of the mechanism. This depends on the $B_2$ charge of the DM particles, $q_{_\rm{DM}}$, on whether the baryonic asymmetry is inherited to the MSSM fields before or after the sphaleron freeze-out, and possibly on similar dark-sector effects.

In the context of pangenesis, the LSP, the generic susy DM candidate, can be rendered unstable by breaking $R$-parity while maintaining baryon triality to ensure proton stability. Even in the presence of R-parity, pangenesis can be considered complementary to models in which the LSP is mostly higgsino or wino and its thermal relic density is not sufficient to account for the entirety of DM. This may also be the case if the LSP belongs to the dark sector, annihilates sufficiently strongly via a dark force and/or is sufficiently light.

\section{Direct Detection of Dark Matter and Collider Signals}
\label{sec:signatures}

Models with gauged $B-L$ provide enhanced prospects for DM direct detection and collider signals. 
The spin-independent scattering cross section of $B_2$-charged DM, per nucleon, via $Z'_{_{B-L}}$ exchange, can be as high as
\beq
\s_{_{B-L}}^\rm{SI}   \!\!  \approx  \pare{3 \times 10^{-44} \cm^2}  q_{_\rm{DM}}^2
\pare{ \frac{g_{_{B-L}}}{0.1} }^4 \!
\pare{ \frac{0.7 \TeV}{M_{_{B-L}}} }^4,
\label{eq:sigmaDD_B-L}
\eeq
where $q_{_\rm{DM}} \sim  \cal{O}(1)$, $g_{_{B-L}}$ is the gauge coupling and $M_{_{B-L}}$ the mass of $Z'_{_{B-L}}$, and we used $m_\rm{DM} = 5 \GeV$.

Collider experiments can probe the existence of a $Z_{_{B-L}}'$ gauge boson. If its invisible decay width exceeds what can be accounted for by neutrinos, such a discovery would point towards a $(B-L)$-charged dark sector.

If the symmetric part of DM annihilates via a dark abelian force, the possible kinetic mixing with the SM hypercharge, $-(\e/2) F_{_Y}^{\m\n}  F'_{_D \: \m\n}$, provides a DM direct detection channel.
For efficient annihilation, the dark $U(1)'_{_D}$ coupling has to be sufficiently large:
if DM is lighter than the $Z'_{_D}$ boson, then $g_{_D}^2 / M_{_D}^2  >  10^{-4} \GeV^{-2} ( 5 \GeV / m_\rm{DM} )$; 
if DM is heavier than the $Z'_{_D}$ boson, then $g_{_D} > 0.1 ( m_\rm{DM} / 5 \GeV )^{1/2}$~\cite{Graesser:2011wi}.
The kinetic mixing is constrained to be roughly $\e \lesssim 10^{-3}$ for $M_{_D} \gtrsim 0.1 \GeV$~\cite{Bjorken:2009mm}.
The spin-independent scattering cross section per nucleon, via $Z'_{_D}$ exchange, is
\beq
\s_{_D}^\rm{SI}    \approx   \pare{10^{-40} \cm^2} 
\pare{ \frac{\e}{10^{-4}} }^2      
\pare{ \frac{g_{_D}}{0.1} }^2     
\pare{ \frac{1 \GeV}{M_{_D}} }^4 .
\label{eq:sigmaDD_dark}
\eeq
This can account for the light DM regions favored by DAMA and CoGeNT~\cite{Savage:2008er,Aalseth:2010vx}. However, the above estimate can vary by several orders, and comfortably satisfy constraints from XENON100~\cite{Aprile:2011hi} and CRESST~\cite{Angloher:2002in}.

The kinetic mixing of a $U(1)'_{_D}$ dark force with $U(1)_{_Y}$ can be probed in current fixed target experiments~\cite{Bjorken:2009mm}.

\section{Conclusions}
\label{sec:conc}
The mechanism of \emph{pangenesis} employs the Affleck-Dine scenario to explain the observed similarities of the visible and dark matter densities in the context of a universe with a generalized $B$ or $B-L$ number equal to zero. We examined one explicit realization, but expect many variations to be possible. Besides supersymmetry, which is typically necessary for the AD mechanism, evidence in favor of pangenesis would include a DM particle of mass $\mathcal{O}(10)$ GeV (already favored by DAMA and CoGeNT), and the likely existence of a $Z'$ boson coupling to standard $B-L$ in the visible sector but with an invisible width driven by decays into dark sector species. The necessity to annihilate the symmetric part of the DM also motivates the presence of a dark $U(1)'$ gauge interaction that kinetically mixes with the SM hypercharge. Both the $B-L$ and the dark $U(1)'$ gauge forces enhance the prospects for DM direct detection.

\section*{Acknowledgements}
We thank Alex Kusenko for reading the manuscript and for useful comments.
We also thank Robert Foot, Michael Graesser, Archil Kobakhidze, Nick Setzer and Luca Vecchi for helpful discussions. This work was supported, in part, by the Australian Research Council.

\bibliography{Bibliography}

\end{document}